\documentclass[aps,prb,reprint,groupedaddress]{revtex4-1}
\usepackage[dvips]{graphicx}
\usepackage{comment}
\usepackage{amsmath}
\usepackage{tabularx}
\usepackage{gensymb}

\begin{document}


\title{Quantum oscillations with non-zero Berry phase from a complex three dimensional Fermi surface in Bi$_2$Te$_3$}


\author{Sourabh Barua}
\email[]{sbarua@iitk.ac.in}
\affiliation{Department of Physics, Indian Institute of Technology Kanpur, Kanpur 208016, India}

\author{K. P. Rajeev}
\affiliation{Department of Physics, Indian Institute of Technology Kanpur, Kanpur 208016, India}


\date{\today}

\begin{abstract}
We performed angle dependent magnetoresistance study of a metallic single crystal sample of Bi$_2$Te$_3$. We find that the magnetoresistance is highly asymmetric in positive and negative magnetic fields for small angles between the magnetic field and the direction perpendicular to the plane of the sample. The magnetoresistance becomes symmetric as the angle approaches 90\degree. The quantum Shubnikov de-Haas oscillations are symmetric and show signatures of topological surface states with Dirac dispersion in the form of non-zero Berry phase. However, the angular dependence of these oscillations suggests a complex three dimensional Fermi surface as the source of these oscillations, which does not exactly conform with the six ellipsoidal model of the Fermi surface of Bi$_2$Te$_3$. We attribute the asymmetry in the magnetoresistance to a mixing of the Hall voltage in the longitudinal resistance due to the comparable magnitude of the Hall and longitudinal resistance in our samples. This provides a clue to understanding the asymmetric magnetoresistance often seen in this and similar materials. Moreover, the asymmetric nature evolves with exposure to atmosphere and thermal cycling, which we believe is either due to exposure to atmosphere or thermal cycling, or both affecting the carrier concentration and hence the Hall signal in these samples. However, the quantum oscillations seem to be robust against these factors which suggests that the two have different origins.
\end{abstract}


\maketitle

\section{Introduction}
Topological insulator materials have attracted a great deal of attention because of their special properties like an insulating bulk with topologically guaranteed conducting surfaces and spin momentum locking of the charge carriers of these surface states, which in turn have a Dirac dispersion.\cite{Ando1,Hasan} However, realising the full potential of these materials in electrical transport measurements has been tough, as the surface conduction is always overwhelmed by a conducting bulk due to defects in the crystal and the surface conduction is at best only a fraction of the total conduction.\cite{Ando1,Hasan,Qi,Barua1} Thus till now, experimentalists have had to rely on indirect methods to filter out the surface contribution in the electrical transport, like the quantum Shubnikov-de Haas oscillations (SdH) \cite{Lahoud,Taskin1,Ren,Qu2010} or the multiple band model for the Hall data.\cite{Taskin2,Xia,Ren} In this report, we too have used the quantum SdH oscillations to shed more light on the signatures of the topological surface states in the electrical transport of Bi$_2$Te$_3$.

Quantum oscillations have by far played the most important role in detecting the surface states in electrical transport measurements. Quantum oscillations occur in the presence of a magnetic field because the energy states in the plane perpendicular to the field direction are quantized into Landau levels. As the field increases, these Landau levels expand in the reciprocal lattice space, leading to periodic crossings of the Fermi surface by successive Landau levels. At these crossings, there is a maximum in the density of states at the Fermi energy and this is manifested as periodic oscillations in the physical properties of the system like resistivity, magnetization etc. \cite{Kittel,Eto} Only the area of the extremal cross section of the Fermi surface perpendicular to the field direction determines the periodicity of these oscillations, as the overlap with the Landau levels is the largest at these places on the Fermi surface. Hence, by rotating the field direction with respect to the crystal axes, information regarding the shape of the Fermi surface can be obtained. This technique has been used by experimentalists to single out the quantum oscillations originating from the surface states in topological insulator materials, as the expected angular dependence for the Fermi surfaces of the surface states and bulk are different. In case of the surface states the Fermi surface is 2D and one expects a $1/\text{\text{cos}}(\theta)$ dependence of the oscillation frequency on the angle $\theta$, which the field makes with the normal to the plane of the surface states. \cite{Yan1,Ren,Taskin2,Lahoud,Qu-F} Quantum oscillations from a bulk Fermi surface, which is reported to be an ellipsoid in these materials, have also been observed and the frequency of oscillations in this case deviates from a $1/\text{cos}(\theta)$ dependence. \cite{Eto,Analytis1,Lahoud,Qu-F} The Fermi surface in both n and p-type Bi$_2$Te$_3$ is believed to consist of six ellipsoids with a non-parabolic dispersion, with two such ellipsoids located in each of the three mirror planes in the Brillouin zone.\cite{Mallinson1968,Kohler1976a,Kohler1976b} The ellipsoids are further tilted in the mirror plane which contains the trigonal and bisectrix axis and only the binary axis has a principal axis of the ellipsoid parallel to it. The conclusion that the dispersion was non-parabolic, was drawn on the basis of the fact that the effective mass of the charge carriers was not a constant for different carrier concentrations and increased with energy. The non-parabolic nature became more strong for Fermi energies above 20 meV and carrier concentration $>$ 1 $\times$ 10$^{-19}$ cm$^{-3}$.\cite{Kohler1976a} The Berry phase is a further fundamental difference between the oscillations from the topological surface states and those from the bulk. It can be extracted from the quantum oscillations and is equal to $\pi$ for the surface states and zero for the bulk, because the surface states have a Dirac dispersion whereas the bulk states are normally expected to have a parabolic dispersion.\cite{Tian,Tang,Qu2010}

Another universal feature of the topological insulators, especially Bi$_2$Se$_3$ and Bi$_2$Te$_3$, is that the magnetoresistance varies linearly with field while classically a initial parabolic rise followed by saturation is expected.\cite{Tang,He2,Yan2,Wang,He1,Zhang,Assaf} While the reason behind this is still not fully understood, there is evidence of the 2D nature of this magnetoresistance, with its magnitude decreasing as the field is tilted from a perpendicular to a parallel direction with respect to the plane of the sample. \cite{Tang,He2,Yan2} This has led to a connection being drawn between the 2D topological surface states and the 2D linear magnetoresistance, as the phenomenon is explained by Abrikosov's theory of quantum linear magnetoresistance, which requires the presence of gapless states with linear dispersion and these are already present in topological insulators.\cite{Wang,He1} Alternatively, it is also postulated that the surface states can lead to linear magnetoresitance by a different mechanism of weak anti-localization (WAL) as explained by the Hikami-Larkin-Nagaoka (HLN) model. \cite{Zhang,Assaf} Besides this there is also a third possibility of inhomogeneity or mobility disorder leading to a linear magnetoresistance as suggested by the Parish-Littlewood model. \cite{He2,Yan2}

Yet another subtle feature of these materials which has not got much attention is that in studies of the angle dependence of the linear magnetoresistance in  Bi$_2$Se$_3$\cite{He2} and Bi$_2$Te$_3$\cite{Yue}, asymmetry in the magnetoresistance for positive and negative fields has been reported. In this report, we point out that such an asymmetry in these materials could arise due to the mixing of the Hall and resistance signals and this maybe difficult to remove because of the large Hall signal in these materials. Our angle dependent magnetoresistance study of a metallic single crystal of Bi$_2$Te$_3$, further shows that the quantum oscillations have a three dimensional origin although they have a Berry phase nearly equal to $\pi$. The asymmetry in magnetoresistance also shows a time dependence from which we infer that the exposure to air may lead to change in the carrier concentration, however this does not seem to affect the quantum oscillations, making us believe that they involve a different set of charge carriers.

\begin{figure}[h!]
\includegraphics[width = 0.5\textwidth]{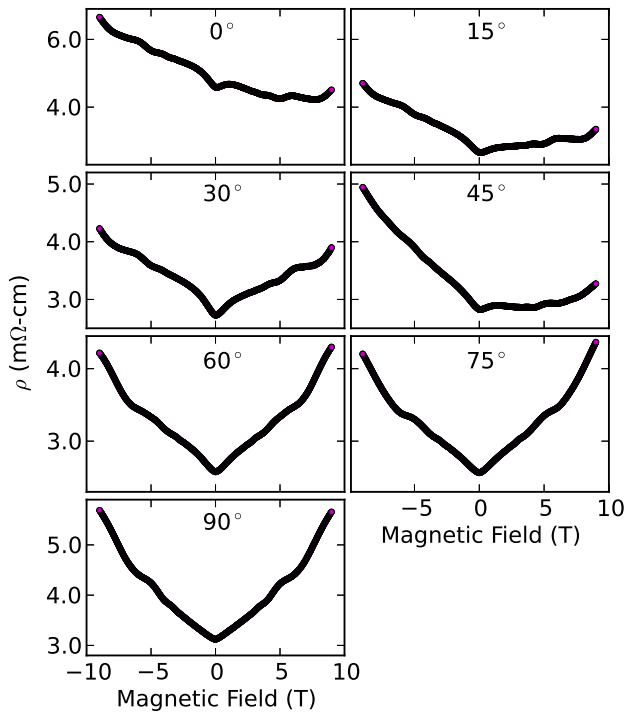}
\caption{Magnetoresistance at different angles measured at 1.5 K showing the linear magnetoresistance and quantum oscillations.}  \label{Fig1}
\end{figure}

\section{Experimental Details}

The single crystal sample of Bi$_2$Te$_3$ studied in this report is the same as the sample number 2 reported in one of our earlier works. \cite{Barua2} The samples were cleaved from a large piece of natural single crystal of Bi$_2$Te$_3$. The X-ray diffraction pattern and energy dispersive X-ray spectra of the sample along with the temperature profile of resistivity and Hall measurement data are shown in that report. The resistivity versus temperature showed that the sample was metallic in nature and the Hall data showed that the sample was p-type. The angle dependent magnetoresistance measurements were done after mounting the sample on a special homemade sample holder with provisions for making measurements at seven different angles from $0\degree$ to $90\degree$ at intervals of $15\degree$. We used an ICEOxford make closed cycle refrigerator $^{DRY}$ICE$^{VTI}$ for the magnetoresistance measurements which can attain a minimum temperature of 1.5 K and provide a magnetic field upto 9 T. Four probe contacts were made using gold wires and silver paint for the electrical transport measurements. The magnetoresistance and Hall measurements reported here for the various angles were all done at 1.5 K. The sample had to be brought to room temperature and exposed to atmosphere for changing the orientation of the sample holder before making a measurement at a new angle.

\section{Results}

\begin{figure}[h!]
\includegraphics[width = 0.5\textwidth]{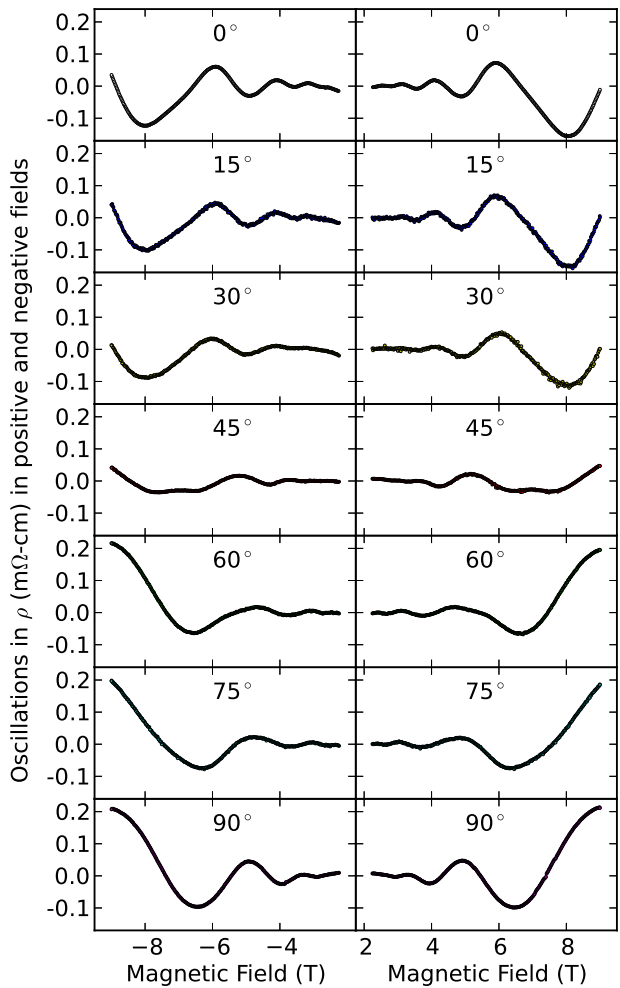}
\caption{Quantum oscillations extracted by subtracting only the polynomial part of the fit to the magnetoresistance, shown for both positive and negative fields at different angles.}  \label{Fig2}
\end{figure}

To study the angle dependence of the magnetoresistance, the magnetoresistance was measured with the magnetic field making different angles with the $c$-axis of the sample, which is perpendicular to the plane of the sample. In all, measurements were performed at seven different angles of $0\degree$, $15\degree$, $30\degree$, $45\degree$, $60\degree$, $75\degree$ and $90\degree$ between the direction of the applied magnetic field and the direction perpendicular to the plane of the sample. In figure \ref{Fig1} we show the magnetoresistance at the different angles. The main feature of the magnetoresistance at all the angles is that it tends to vary linearly with magnetic field and also shows quantum Shubnikov-de Haas (SdH) oscillations. The quantum oscillations are present in both positive and negative fields and so is the linear magnetoresistance. However, the magnetoresistance is highly asymmetric in positive and negative fields for the small angles $0\degree$, $15\degree$, $30\degree$ and $45\degree$ and becomes symmetric as the angle increases. The measurements at different angles were performed in the order $0\degree$, $90\degree$, $45\degree$, $75\degree$, $60\degree$, $30\degree$ and $15\degree$. We note from figure \ref{Fig1}, that the resistance initially drops as seen by comparing the measurements at $0\degree$, $90\degree$ and $45\degree$ and then more or less stabilizes towards the last measurements. Furthermore, the asymmetry also decreases with successive measurements, which is the reason why the $45\degree$ curve is more asymmetric than the $30\degree$ curve even though the general trend is a decrease in the asymmetry with increasing angle. We discuss this in more detail in the discussion section.

We fitted the magnetoresistance at the different angles using an expression for the quantum oscillations along with a polynomial background magnetoresistance.\cite{Tang,Barua2,Schneider} The expression used for the fitting is given below.
\begin{equation}
\label{eqn1}
\begin{split}
\rho = A \text{exp}(-\pi/\mu B)B^{1/2}\text{cos}[2\pi(B_\text{F}/B+\beta+0.5)]\\+\alpha+\gamma B+\delta B^2
\end{split}
\end{equation}

Here, $B$ is the absolute value of the magnetic field, $A$ is the amplitude, $\mu$ is the mobility of the charge carriers, $B_\text{F}$ is the frequency of the quantum oscillations and $2\pi\beta$ is the Berry phase. $\alpha$, $\gamma$ and $\delta$ are the coefficients of the polynomial background. The fit was done for fields above 2.25 T, for both positive and negative fields. For the $45\degree$ measurement we fitted only a polynomial background without the expression for the quantum oscillations as we observed that a fit to quantum oscillations along with the polynomial part was not good. We discuss this later in the paper. The quantum oscillations appear symmetric in the positive and negative magnetic fields as shown in figure \ref{Fig1}. To confirm this we extracted the quantum oscillations by subtracting only the polynomial part of the fit from the actual data and the results are shown in figure \ref{Fig2}. It is clear from the figure that the quantum oscillations are symmetric in positive and negative fields for all the angles. However, the oscillations for the smaller angles are not perfectly symmetric because of the different background magnetoresistance that has to be fitted in these cases. The oscillation for the $45\degree$ angle clearly indicates that the oscillations are not simple but could be a superposition of two different frequencies. We discuss this in more detail in the next section.

\begin{figure}[h!]
\includegraphics[width=0.5\textwidth]{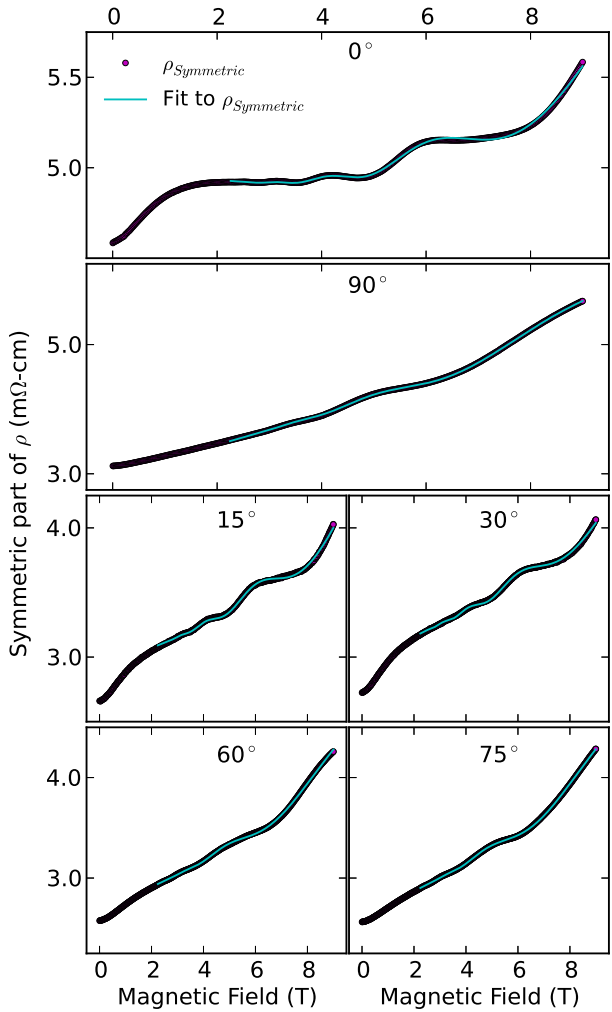}
\caption{Symmetric part of the magnetoresistance along with fit to the symmetric part at various angles. The fit consists of the expression for the quantum oscillations alongwith a parabolic part.}
\label{Fig3}
\end{figure}

\section{Discussion}

\begin{figure}[h!]
\includegraphics[width = 0.5\textwidth]{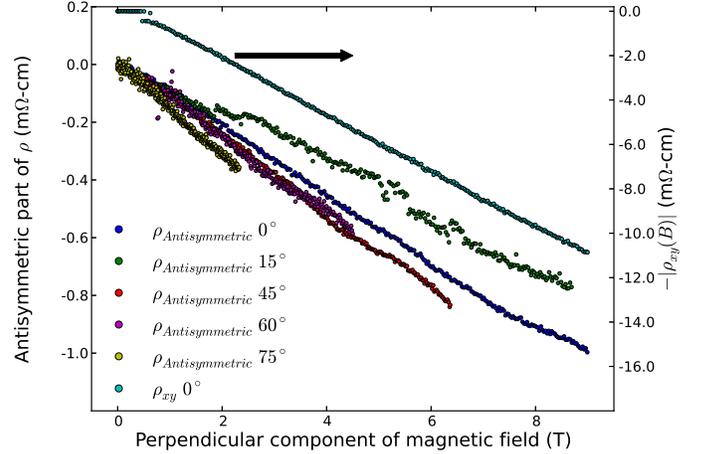}
\caption{Comparison of antisymmetric part of $\rho(B)$ at different angles with -$\left|\rho_{xy}(B)\right|$ at $0\degree$which was done alongwith the magnetoresistance. The antisymmetric part are plotted on the left while the Hall resisitivity is plotted on the right.}  \label{Fig5}
\end{figure}

To understand the reason behind the asymmetry in the magnetoresistance in the positive and negative fields, we calculated the symmetric and antisymmetric parts for the magnetoresistance at the different angles by calculating $\frac{\rho(B)+\rho(-B)}{2}$ for the symmetric part and  $\frac{\rho(B)-\rho(-B)}{2}$ for the antisymmetric part, where $\rho(B)$ is the resistivity at field $B$. The results for the symmetric part of the magnetoresistance are shown in figure \ref{Fig3}. For the antisymmetric part of the magnetoresistance, we felt that it could be coming from the Hall voltage because it varied linearly with field and its magnitude decreased as the angle increased. The decrease in magnitude of the antisymmetric part at higher angles is expected if it is coming from a Hall voltage, because the Hall signal is proportional to the perpendicular component of the magnetic field. To check whether the antisymmetric part could be arising from a Hall signal, we repeated the magnetoresistance measurements at different angles, however this time they were done alongwith simultaneous Hall measurements. In figure \ref{Fig5}, we show the antisymmetric part calculated for these set of magnetoresistance at different angles and compare it with the actual Hall resistivity done simultaneously with the magnetoresistance at $0\degree$. The antisymmetric part at the different angles have been plotted against effective field $B\text{cos}\theta$ at each angle of measurement $\theta$. From the figure it is clear that the antisymmetric part of magnetoresistance at different angles falls, more or less, on the same straight line when plotted against effective magnetic field and further the Hall resistivity is larger than the antisymmetric part of the magnetoresistance, implying that the latter could be arising from the Hall signal. Thus, the mixing of the Hall voltage in the longitudinal resistivity of our sample could be because of the fact that the Hall resistivity in our sample is larger than the longitudinal resistivity, which would lead to such mixing even for very small mismatch in the alignment of the four probe contacts that are placed in a straight line on the sample. In fact, similar situation has been observed in other topological insulator materials also and in some of those reports the magnetoresistance data was symmetrized to extract the Hall and longitudinal resistivity.\cite{Pan,Analytis2,Petrushevsky,Klijnsma,KimHJ} This could also be the reason behind the asymmetry seen in the magnetoresistance for positive and negative fields while studying the angular dependence of the magnetoresistance in Bi$_2$Te$_3$ and Bi$_2$Se$_3$.\cite{Yue,He2}

\begin{table}[h!]
\caption{Table of parameters obtained from fit to symmetric part of magnetoresistance for different angles alongwith the uncertainties in their determination. $B_\text{F}$ is the frequency of quantum oscillations, $\beta$ is the Berry phase factor and $\mu$ is the mobility of the charge carriers.
\label{Table1}}
\begin{center}
\begin{tabularx}{0.48\textwidth}{|X|X|X|X|}\hline
Angle &  $B_\text{F}$ (Tesla$^{-1}$) & $\beta$ & $\mu$ ($\text{cm}^2/\text{V-s}$)\\ \hline
0\degree  & 12.55 $\pm$ 0.07 & 0.366 $\pm$ 0.009& 0.176 $\pm$ 0.002\\ \hline
15\degree & 12.64 $\pm$ 0.03 & 0.352 $\pm$ 0.004& 0.145 $\pm$ 0.001\\ \hline
30\degree & 12.92 $\pm$ 0.06 & 0.321 $\pm$ 0.009& 0.153 $\pm$ 0.002 \\ \hline
60\degree & 10.04 $\pm$ 0.09 & 0.409 $\pm$ 0.014& 0.164 $\pm$ 0.002\\ \hline
75\degree & 8.39  $\pm$ 0.07 & 0.654 $\pm$ 0.011& 0.178 $\pm$ 0.001\\ \hline
90\degree & 10.35 $\pm$ 0.04 & 0.351 $\pm$ 0.005& 0.273 $\pm$ 0.004\\ \hline
\end{tabularx}
\end{center}
\end{table}

\begin{figure}[h!]
\includegraphics[width = 0.5\textwidth]{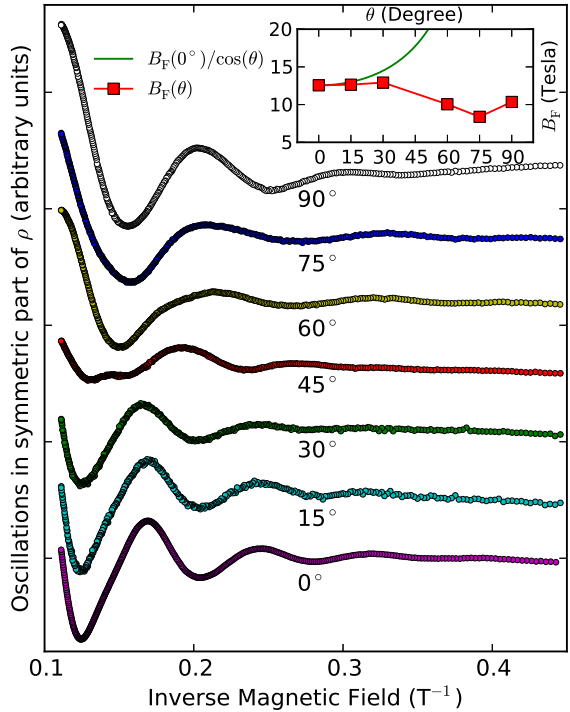}
\caption{Quantum oscillations extracted from the symmetric part of the magnetoresistivity by subtracting the polynomial part of the fit. The oscillations at different angles are offset for clarity. The inset shows the $B_\text{F}$ values for the oscillations at different angles extracted from the SdH fit. For the $45\degree$ angle, we could not fit the SdH oscillation with one frequency and hence it is not shown. The values deviate from a $B_\text{F}(0\degree)/\text{cos}(\theta)$ dependence which is expected from a two dimensional Fermi surface.}  \label{Fig6}
\end{figure}

In figure \ref{Fig3} we have also fitted the symmetric part of the magnetoresistance with equation \ref{eqn1} which consists of the expression for the quantum oscillations alongwith a parabolic background fit. The parameters obtained from the fit are given in table \ref{Table1} alongwith the uncertainty in their estimates. The berry phase factor for the different angles is near to the value of 0.5 expected for Dirac charge carriers of the topological surface states. We can further extract the quantum oscillations in the symmetric part by subtracting only the parabolic background. The quantum oscillations so extracted for the different angles are shown in figure \ref{Fig6} with respect to inverse magnetic field, in which these oscillations are periodic. For the $45\degree$ measurement, we only fitted the parabolic part to the magnetoresistance and then subtracted it from the actual data to get the oscillations. From the figure it is clearly visible, that the oscillations at $0\degree$, $15\degree$ and $30\degree$ are almost similar and form one set of oscillations and those at $60\degree$, $75\degree$ and $90\degree$ are similar and form another set of oscillations. This is also reflected in the fact that the frequencies of the oscillations in each set is more or less equal as can be seen from the frequency values ($B_\text{F}$) at each angle obtained from the fit and given in table \ref{Table1} and the inset of figure \ref{Fig6}. The angular dependence of the oscillation frequency does not follow a $1/\text{cos}\theta$ dependence expected for a two dimensional Fermi surface where $\theta$ is the angle between the magnetic field direction and the $c$-axis of the sample. This is seen in the inset of figure \ref{Fig6}. Instead, the angular dependence of the oscillation frequency of our sample shows a behaviour which might come from a complex three dimensional Fermi surface. This is counterintuitive to the non-zero berry phase obtained from the fit for these oscillations, as oscillations with Berry phase of $\pi$ are supposed to arise from topological surface states which have a 2D Fermi surface.

The three dimensional nature of the Fermi surface behind the quantum oscillations in our sample as revealed from the angle dependence in the inset of figure \ref{Fig6}, merits some attention because it is different from the angular dependence for a three dimensional Fermi surface seen in these materials.\cite{Eto,Analytis1,Lahoud,Qu-F,Kohler1976a} In these reports the frequency increases as the magnetic field is inclined from the $c$-axis or trigonal axis. As mentioned earlier, the Fermi surface in p-type Bi$_2$Te$_3$ is supposed to consist of six ellipsoids in the mirror planes. The location of the Fermi surfaces with respect to the three axes, trigonal, binary and bisectrix was such that only one oscillation frequency was expected for a field parallel to the trigonal axis while two different frequencies were expected for a field parallel to the binary and bisectrix axes.\cite{Kohler1976a,Kohler1976b,Mallinson1968} Even then only one frequency was observed in fields parallel to the binary and bisectrix axis respectively till K\"{o}hler discovered a second frequency in fields parallel to the binary axis.\cite{Kohler1976a} However, it was common to observe harmonics in the SdH oscillations due to spin-splitting in case of fields parallel to all the axes.\cite{Kohler1976a,Kohler1976b,Kulbachinskii1992} It is to be noted that in p-type Bi$_2$Te$_3$, for higher carrier concentrations, two frequencies could be observed in SdH oscillations even for field parallel to the trigonal axis because of a second valence band.\cite{Kohler1976a,Middendorff1972} In our case, unlike the other reports, the frequency decreases as we incline the magnetic field away from the $c$-axis. We also do not see more than one frequency either along the $c$-axis or when the field is perpendicular to it, which is not uncommon as discussed earlier. It is to be noted that when the field is perpendicular to the $c$-axis, we do not know the exact alignment of the field with respect to the crystal axes and we only know that the field is in the plane containing the binary and bisectrix axis. However, the interesting part is that there are more than one frequency in the oscillations at 45\degree. It is clear from figure \ref{Fig6} that the oscillations at $45\degree$ are a superposition of more than one frequency which are not harmonics and could be the result of the interference of the 0\degree and 90\degree oscillations. Thus the three dimensional Fermi surface in our sample seems to be different and the six ellipsoidal Fermi surface model of Bi$_2$Te$_3$ may not be able to account for it.

Another aspect of the magnetoresistance of our sample is that the asymmetry decreased with each successive measurement at a different angle and the resistivity also dropped. Between each such measurement the sample was thermally cycled and also exposed to atmosphere because the sample had to be brought to room temperature and taken out of the cryostat for changing the angle. Both of the above changes can be explained if we assume that there is an increase in the charge carrier concentration of the sample on thermal cycling or exposure to air as this will lead to a decrease in the Hall voltage which in turn will decrease the asymmetry. The resistivity which is inversely proportional to the charge carrier concentration will also decrease. This has been the case in Bi$_2$Se$_3$, where exposure to air reduced the resistivity and led to increase in carrier concentration.\cite{Analytis2} It has also been seen in Bi$_2$Se$_3$ and Bi$_2$Te$_3$, that exposure to atmosphere generally leads to a drop in resistance and n-type doping\cite{Analytis1,Analytis2,Hoefer2014}, however exposure to atmosphere can lead to both n-type as well as p-type doping depending upon which constituent of air has the predominant effect.\cite{Brahlek} However the quantum oscillations in our sample seem to be immune to the thermal cycling and exposure to air of the sample, as the oscillations for several successive measurements done at the same angle superpose on one another even when there is considerable change in the background magnetoresistance, which in turn is a reflection of Hall coefficient change. This would suggest that the charge carriers involved in the quantum oscillations are different from those involved in the Hall resistivity. This would be the case if the quantum oscillations were coming from the surface states, as indicated by the non-zero berry phase in our sample and if the Hall signal were coming from the bulk. However the three dimensional nature inferred from the angle dependence of the quantum oscillations makes this justification difficult. Even so, it must be pointed out that the immunity of the quantum oscillations in our sample is similar to the case of Bi$_2$Se$_3$, where it was shown that the bulk carrier concentration increased and bulk carrier mobility decreased on exposure to air while the carrier concentration and carrier mobility of the topological surface states were largely unaffected.\cite{Brahlek} The effect of the exposure on the surface states is however not clear, as some report washing away of surface oscillations on exposure while others report robustness of the surface states against exposure as evident from their non-linear Hall data.\cite{Brahlek,Analytis2} The surface oscillations could be washed away because exposure induced doping moves the Fermi level into the bulk band because of which the bulk dominates over the surface in transport measurements. Thus the effect of exposure to atmosphere on the surface states is still to be understood fully.

Finally, it must be noted that the non-parabolic dispersion of the highest valence band in p-type Bi$_2$Te$_3$ as discovered by K\"{o}hler could be because of the combination of the parabolic dispersion of the bulk states and linear dispersion of the surface states. Alternatively, it may be that the non-zero Berry phase in our sample has some connection with the non-parabolic bulk valence bands in Bi$_2$Te$_3$.

\section{Conclusion} In conclusion, we have observed that the quantum oscillations in metallic single crystals of Bi$_2$Te$_3$ show a three dimensional nature although they have a Berry phase close to that expected for the surface states of topological insulator materials. The three dimensional behaviour of the Fermi surface also does not conform to the six ellipsoidal Fermi surface model of Bi$_2$Te$_3$. The asymmetry seen in the magnetoresistance of these materials has been explained to be arising from the mixing of the large Hall signal in these materials with the longitudinal resistivity. The Hall signal seems to be affected by thermal cycling and exposure to atmosphere of these samples as inferred from the evolution of the asymmetric part of the magnetoresistance with the repeated measurements, however the quantum oscillations are unaffected leading us to believe that these two have different origins.

\begin{acknowledgments}
SB would like to acknowledge CSIR, India for financial support in the form of research fellowship.
\end{acknowledgments}

\bibliographystyle{apsrev4-1}
\bibliography{ref}

\end{document}